# Накат нелинейно деформированных волн на берег

*И.И.Диденкулова[1,2], Н.Заибо[3], А.А.Куркин[2], член-корреспондент РАН Б.В.Левин[4,5], Е.Н.Пелиновский[1,2], Т.Соомере[6]*

[1] Отдел нелинейных процессов в геофизике, Институт прикладной физики РАН

[2] Кафедра прикладной математики, Нижегородский государственный технический университет

[3] Кафедра физики, Университет Антильских островов, Гваделупа, Франция

[4] Институт океанологии РАН, Москва

[5] Институт морской геологии и геофизики ДВО РАН, Южно-Сахалинск

[6] Институт кибернетики, Таллиннский технологический университет, Таллинн, Эстония

February 23, 2006.

Проблема наката морских волн на берег обсуждается в рамках точных решений нелинейной теории длинных необрушенных волн. Новым моментом здесь является учет анизотропии волн на мелководье, когда крутизна переднего склона волны превышает крутизну заднего склона. Показано, что высота наката волн на берег сильно возрастает с увеличением крутизны переднего склона, в то время как глубина отката сравнительно мало от нее зависит. Эти результаты объясняют, почему волны цунами с крутым фронтом, как это было во время цунами 2004 года в Индийском океане, проникают на мелководье значительно дальше, чем волны с симметричным профилем.

# Runup of nonlinear deformed waves on a beach

*Didenkulova I., Zahibo N., Kurkin A., Levin B., Pelinovsky E., and Soomere T.*

The problem of the sea wave runup on a beach is discussed in the framework of the rigorous solutions of the nonlinear shallow-water theory. Key and novel moment here is the accounting of the asymmetric waves when the face slope steepness exceeds the back slope steepness. It is shown that the runup height growth with increase of the face slope steepness meanwhile the rundown characteristics are weakly depend from the face slope steepness. These results explain why the tsunami waves with steep front (as during the 2004 tsunami in the Indian Ocean) penetrate inland on large distances to compare with wave with symmetrical shape.

Проблема наката длинных необрушенных волн на плоский откос является достаточно хорошо разработанной с математической точки зрения в рамках нелинейной теории мелкой воды, допускающей аналитическое решение с помощью преобразования Карриера-Гринспана (Carrier & Greenspan, 1958). Разнообразные примеры подходящих волн рассмотрены в литературе; обзор старых работ дан в книге (Пелиновский, 1996), приведем также последние публикации (Massel & Pelinovsky, 2001; Carrier et al, 2003; Kânoğlu, 2004; Tinti & Tonini, 2005). Однако во всех упомянутых работах подходящая волна была симметричной или антисимметричной с одинаковой крутизной переднего и заднего склонов. В результате, формулы для высоты наката могут быть параметризованы, и в них входят высота и длина подходящей волны, а также расстояние до берега. Между тем, как показывают многочисленные наблюдения, сделанные во время цунами 2004 года в Индийском океане, к берегу подходит сильно деформированная волна с заметной крутизной переднего склона. Здесь мы покажем, что волна с увеличенной крутизной переднего склона проникает на побережье дальше, чем волна с симметричным профилем.

Как известно, длина волны цунами достаточно велика, поэтому адекватной моделью для описания наката волн цунами на берег является нелинейная теория мелкой воды. Считая жидкость идеальной, а волну распространяющейся по нормали к берегу, запишем основные уравнения мелкой воды

$$\frac{\partial u}{\partial t} + u\frac{\partial u}{\partial x} + g\frac{\partial \eta}{\partial x} = 0, \qquad \frac{\partial \eta}{\partial t} + \frac{\partial}{\partial x}\left([h(x)+\eta]u\right) = 0, \qquad (1)$$

где $\eta$ – возвышение водной поверхности, $u$ – горизонтальная скорость водного потока, $g$ – ускорение силы тяжести и $h(x)$ – невозмущенная глубина бассейна. Геометрия задачи показана на рис. 1: зона наката протяженности $L$ (угол наклона берега постоянен и равен $\alpha$) сопряжена с участком ровного дна. Мы будем предполагать, что волна подходит к берегу справа, и ее форма задана в точке $X + L$ (такая ситуация реализуется обычно при лабораторном моделировании движения волны от волнопродуктора).

Рассмотрим сначала движение волны на ровном участке ($x > L$). В этом случае уравнения мелкой воды допускают точное решение в виде простой (римановой) волны (см., например, Вольцингер и др., 1989)

$$\eta(x,t) = \eta_0\left(t + \frac{x - X - L}{V(\eta)}\right), \qquad \text{or} \qquad t + \frac{x - X - L}{V(\eta)} = \tau(\eta), \qquad (2)$$

где $\tau(\eta)$ - обратная функции к $\eta_0(t)$, описывающая форму приходящей из океана волны в точке $X + L$ (или волну, генерируемую волнопродуктором), и характеристическая скорость $V(\eta)$ есть

$$V(\eta) = 3\sqrt{g(h+\eta)} - 2\sqrt{gh}. \qquad (3)$$

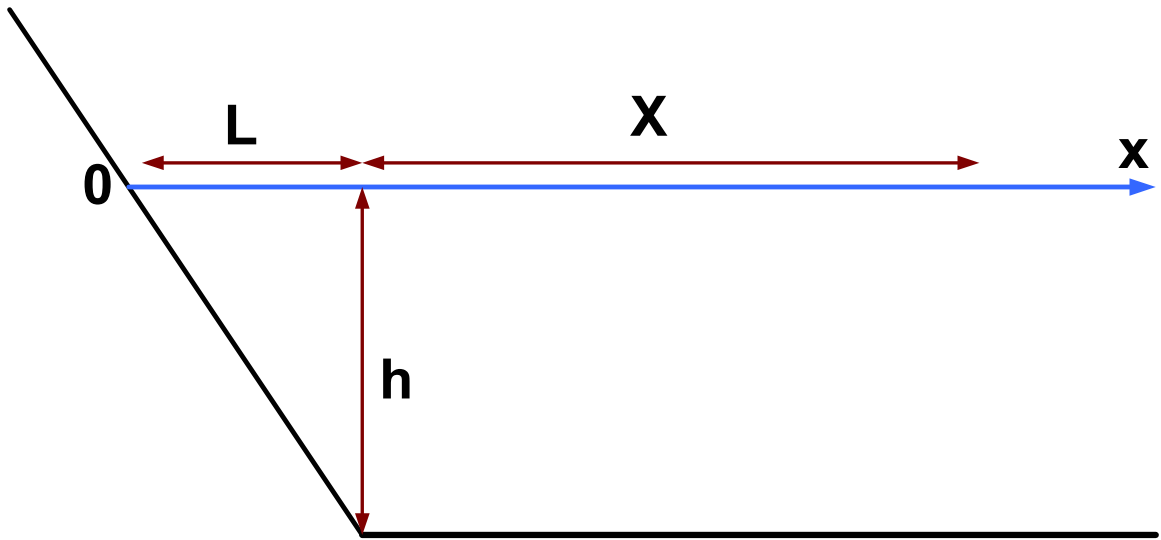

Рис. 1. Геометрия прибрежной зоны

Процесс нелинейной деформации волны на мелководье в рамках (2) хорошо известен и может быть проанализирован для волны произвольной амплитуды, тем не менее, в дальнейшем мы будем использовать приближение слабой нелинейности, когда характеристическая скорость аппроксимируется линейным (по высоте волны) выражением

$$V(\eta) \approx c\left(1 + \frac{3\eta}{2h_0}\right), \qquad c = \sqrt{gh_0}. \qquad (4)$$

С расстоянием крутизна волны возрастает, и ее локальная крутизна ($s = \partial\eta/\partial x$) легко находится из (2)

$$s(x) = \frac{s_0}{1 - \dfrac{X+L-x}{L_n}}, \qquad (5)$$

где $s_0$ - крутизна подходящей к берегу волны и $L_n$ – длина нелинейности, определяющая расстояние, на котором волна опрокидывается. В частности, если к берегу подходит синусоидальная волна $\eta(t) = a\sin(\omega t)$, то волна опрокидывается на расстоянии

$$L_n = \frac{2ch_0}{3\omega a} \qquad (6)$$

при этом ее крутизна обращается в бесконечность (начальная крутизна есть $s_0 = a\omega/c$).

Нелинейная деформация волны на мелководье приводит к генерации обертонов, которые могут быть рассчитаны в приближении слабой нелинейности; задачи такого рода активно исследуются в нелинейной акустике (Руденко и Солуян, 1975; Гурбатов и др., 1990)

$$\eta(\tau, x) = \sum_{n=1}^{\infty} A_n(x) \sin\left[n\omega\left(t + \frac{x - X - L}{c}\right)\right], \qquad (7)$$

где амплитуды гармоник зависят от расстояния

$$A_n(x) = 2a \frac{L_n}{n(X + L - x)} J_n\left(\frac{n(X + L - x)}{L_n}\right), \qquad (8)$$

где $J_n$ – функция Бесселя. Более удобно, однако, исключить расстояние и связать спектр волны (8) с ее крутизной (5); эта зависимость представлена на рис. 2

$$A_n(s) = \frac{2a}{n(1 - s_0/s)} J_n\left(n\left[1 - \frac{s_0}{s}\right]\right). \qquad (9)$$

Амплитуды гармоник возрастают с ростом крутизны, стремясь к предельным значениям при большой крутизне, в то время как амплитуда первой гармоники убывает. В результате удается оценивать спектр нелинейно деформируемой волны по измеряемой крутизне волны, и, по существу, исключить из анализа проблемы наката стадию подхода волны к откосу. Волну вида (7) с амплитудами (9) в точке $x = L$ ($t – X/c$ может быть переобозначено как новое время *t*) можно рассматривать как исходную при решении задачи о трансформации и накате волны на плоский откос.

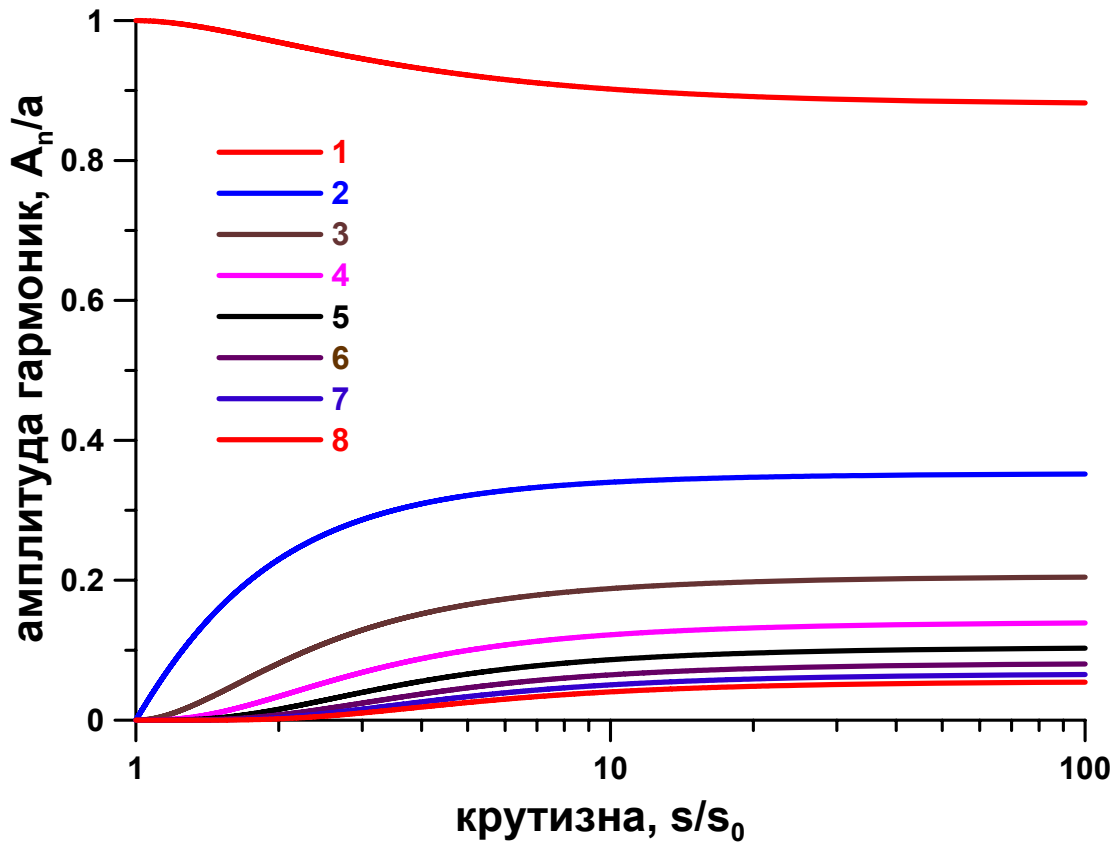

Рис. 2. Связь амплитуд гармоник с крутизной волны

В области $0 < x < L$ необходимо вернуться к решению нелинейных уравнений мелкой воды (1). Оно получается с помощью преобразования Карриера-Гринспана, о чем мы уже упоминали вначале статьи. В частности, если интересоваться только максимальной высотой наката (как и максимальной глубиной отката), то достаточно рассмотреть линейную задачу о колебаниях уровня воды на урезе; этот результат, строгий в математическом отношении, описан в книге Пелиновского (1996). Решение линейной задачи получается достаточно просто и колебания уровня воды на урезе есть

$$R(t) = \eta(0,t) = P \sum_n \sqrt{n} A_n \cos\left(n\omega t + \frac{\pi}{4}\right), \qquad P = 2\pi\sqrt{\frac{2L}{\lambda}}, \qquad (10)$$

где $\lambda$ - длина волны на участке ровного дна, определяемая через заданную частоту волны $\omega$ линейным дисперсионным соотношением ($\lambda = 2\pi c/\omega$). Если волна монохроматическая, то из (10) легко видеть, что максимальная высота волны есть $R_{sin} = aP$ (см., например,

Пелиновский, 1996). Глубина отката в синусоидальной волне равна высоте наката. Дальности наката и отката, естественно, также одинаковы и равны $X_{sin} = R_{sin}/\alpha$. Расчеты максимальной высоты наката и глубины отката (нормированные на высоту наката синусоидальной волны $R_{sin}$) при подходе нелинейно деформируемой волны (7) выполнены численно в рамках (10); результаты расчетов представлены на рис. 3. Как видим, глубина отката слабо зависит от крутизны волны, меняясь не более чем на 30%, и для нее оценки, сделанные по формулам монохроматической волны, являются удовлетворительными. Высота же наката, напротив, сильно зависит от крутизны волны, стремясь к бесконечности для ударной волны (бора) в рамках данной модели (на самом деле, обрушение ограничивает высоту волны на берегу). Эту кривую в первом приближении можно аппроксимировать почти корневой зависимостью

$$R_{\max} = 2\pi a \sqrt{\frac{2L}{\lambda}} \left[\frac{s}{ak}\right]^{0.47}. \qquad (11)$$

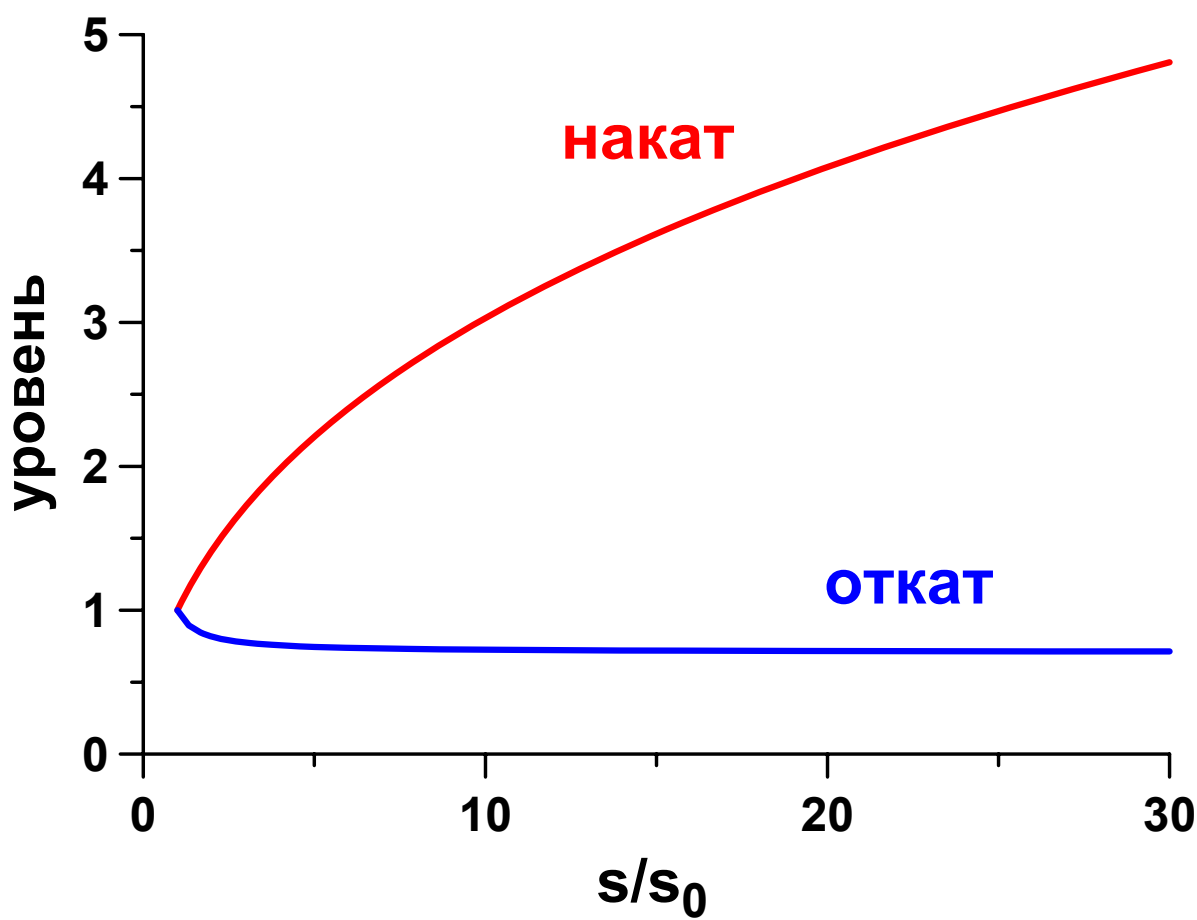

**Рис. 3.** Связь высоты наката и глубины отката с крутизной падающей волны

Ранее при моделировании процесса наката волны цунами на берег неоднократно отмечалось, что высота наката связана с амплитудой подходящей волны нелинейной зависимостью, эти работы цитируются в книге (Пелиновский, 1996). В частности, для солитона длина волны зависит от амплитуды как $\lambda \sim a^{-1/2}$, что дает $R \sim a^{5/4}$. Однако нелинейная зависимость получается для волн по существу любой симметричной формы, хотя для них длительность никак не связана с амплитудой. Качественное объяснение этого эффекта, связанное с нелинейной деформацией волны, было предложено в (Осипенко, Пелиновский, 1992), однако количественное объяснение, как и формулировка определяющих параметров, отсутствовало. В рамках развитой здесь теории становится понятной роль крутизны волны, как определяющего параметра, при расчетах высоты наката волны цунами. Развитая теория предсказывает, что накат обрушенной волны (или, по крайней мере, сильно деформированной волны) может быть значительно более сильным, чем накат волны симметричной формы. Известные случаи наблюдения глубокого проникновения волны с передним обрушенным склоном вглубь побережья (в том числе и во время катастрофического цунами 2004 года в Индийском океане) также могут быть интерпретированы в рамках данной теории.

Зная высоту наката и глубину отката можно рассчитать дальность наката и отката, поскольку угол наклона берега известен. Нелинейная теория позволяет также исследовать характеристики обрушения волн на берегу. В частности, нелинейная деформация волны на подходе к берегу приводит к более раннему обрушению переднего склона. Количественное исследование этого процесса будет дано в более полной статье.

Частично эта работа поддержана грантами РФФИ (05-05-64265), ИНТАС (03-51-4286, 05-1000008-8080) и научной школы Б.В.Левина, и для Т.С. by grant from the Estonian Science Foundation. Estonian Science Foundation. (5762).

## Литература

**Вольцингер Н.Е., Клеванный К.А., Пелиновский Е.Н.** Длинноволновая динамика прибрежной зоны. Л.: Гидрометеоиздат, 1989.

**Гурбатов С.Н., Малахов А.И., Саичев А.И.** *Нелинейные случайные волны в средах без дисперсии.* М.: Наука, 1990.

**Осипенко Н.Н., Пелиновский Е.Н.** Нелинейная трансформация и накат длинных волн на берег. *Океанология*, 1992, т. 32, № 4, 640 - 646.

**Пелиновский Е.Н.** *Гидродинамика волн цунами*. Нижний Новгород: ИПФ РАН, 1996.

**Руденко О., Солуян С.** *Теоретические основы нелинейной акустики.* М., Наука, 1975.

**Carrier G.F., Greenspan H.P.** Water waves of finite amplitude on a sloping beach. *J. Fluid Mech.*, 1958, vol. 4, 97 - 109.

**Carrier G.F., Wu T.T., Yeh H.** Tsunami run-up and draw-down on a plane beach. *J. Fluid Mech.*, 2003, vol. 475, 79-99.

**Kânoğlu U.** Nonlinear evolution and runup-rundown of long waves over a sloping beach. *J. Fluid Mech.*, 2004, vol. 513, 363-372.

**Massel S.R., Pelinovsky E.N.** Run-up of dispersive and breaking waves on beaches. *Oceanologia*, 2001, vol. 43, No. 1, 61 – 97.

**Tinti S., Tonini R.** Analytical evolution of tsunamis induced by near-shore earthquakes on a constant-slope ocean. *J. Fluid Mech.*, 2005, vol. 535, 33-64.